\documentclass[showpacs,aps,prd,reprint,superscriptaddress,nofootinbib,longbibliography]{revtex4-2}

\usepackage[dvipdfmx]{graphicx}
\usepackage{amsmath,amssymb,bm,color,longtable,mathrsfs,amsfonts,slashed,ulem}

\usepackage[colorlinks=true, pdfstartview=FitV, linkcolor=magenta,citecolor=blue, urlcolor=magenta,
bookmarks=true, bookmarksnumbered=true, breaklinks]{hyperref}

\allowdisplaybreaks

\newcommand \beq{\begin{eqnarray}}
\newcommand \eeq{\end{eqnarray}}

\newcommand{\Slash}[1]{{\ooalign{\hfil/\hfil\crcr$#1$}}}
\newcommand{\tr}{{\rm tr}}

\newcommand{\Nc}{N_{\rm c}}
\newcommand{\Nf}{N_{\rm f}}

\newcommand{\lqcd}{\Lambda_{\rm QCD}}
\newcommand{\vp}{ {\bm p}}

\newcommand{\la}{\langle}
\newcommand{\ra}{\rangle}

\newcommand{\calL}{\mathcal{L}}

\newcommand{\calP}{\mathcal{P}}

\newcommand{\rmd}{\mathrm{d}}
\newcommand{\rmi}{\mathrm{i}}
\newcommand{\rme}{\mathrm{e}}

\newcommand{\MS}{ \overline{\rm MS} }

\begin{document}
\begin{flushright}
\end{flushright}

\title{Thermal effects on sound velocity peak and conformality in isospin QCD}

\author{Ryuji Chiba}
\email{ rjchiba@nucl.phys.tohoku.ac.jp }
\affiliation{ Department of Physics, Tohoku University, Sendai 980-8578, Japan}

\author{Toru Kojo}
\email{ toru.kojo.b1@tohoku.ac.jp }
\affiliation{ Department of Physics, Tohoku University, Sendai 980-8578, Japan}

\author{Daiki Suenaga}
\email{daiki.suenaga@kmi.nagoya-u.ac.jp}
\affiliation{Kobayashi-Maskawa Institute for the Origin of Particles and the Universe, Nagoya University, Nagoya, 464-8602, Japan}
\affiliation{Research Center for Nuclear Physics, Osaka University, Ibaraki 567-0048, Japan}

\date{\today}

\begin{abstract}
We study thermal effects on equations of state (EOS) in isospin QCD, utilizing a quark-meson model coupled to a Polyakov loop. 
The quark-meson model is analyzed at one-loop that is the minimal order to include quark substructure constraints on pions which condense at finite isospin density. 
In the previous study we showed that the quark-meson model at zero temperature produces the sound velocity peak 
and the negative trace anomaly in the domain between the chiral effective theory regime at low density and the perturbative QCD regime at high density,
in reasonable agreement with lattice simulations.
We now include thermal effects from quarks in the Polyakov loop background and examine EOS, 
especially the sound velocity and trace anomaly along isentropic trajectories.
At large isospin density, there are three temperature windows;
(i) the pion condensed region with almost vanishing Polyakov loops, (ii) the pion condensed region with finite Polyakov loops, and (iii) the quark gas without pion condensates.
In the domain (i), the gap associated with the pion condensate strongly quenches thermal excitations. 
As the system approaches the domain (ii), thermal quarks, which behave as non-relativistic particles, 
add energy density but little pressure, substantially reducing the sound velocity to the value less than the conformal value 
while increasing the trace anomaly toward the positive value.
Approaching the domain (iii), thermal quarks become more relativistic as pion condensates melt, increasing sound velocity toward the conformal limit.
Corrections from thermal pions are also briefly discussed.

\end{abstract}


\maketitle


\section{Introduction}
\label{sec:Introduction}

Recent observations for two-solar mass neutron stars (NSs) \cite{Arzoumanian:2017puf,Fonseca:2016tux,Demorest:2010bx,Cromartie:2019kug,Fonseca:2021wxt,Antoniadis:2013pzd}
and
the inferred radii of $\sim 12$-13 km \cite{TheLIGOScientific:2017qsa,Miller:2021qha,Riley:2021pdl,Raaijmakers:2021uju},
together with the low density nuclear constraints \cite{Drischler:2021bup,Sun:2023xkg,Togashi:2017mjp},
imply that
the equation of state (EOS) in QCD exhibits rapid soft-to-stiff evolution \cite{Kojo:2020krb}, i.e.,
the pressure grows rapidly as density increases.
A useful measure for the evolution of stiffness is (adiabatic) sound velocity $c_s = ( \partial P/\partial \varepsilon)^{1/2}$ 
where the derivative is taken for a fixed entropy.
In the dilute regime, nuclear theories predict small $c_s^2$ of $\sim 0.1$ 
at baryon density $n_B \simeq n_0$ \cite{Drischler:2021bup} ($n_0\simeq 0.16\, {\rm fm}^{-3}$: nuclear saturation density),
while in the high density limit $c_s^2$ approaches $1/3$. 
In the intermediate  regime, there is a domain where $c_s^2$ makes a peak whose height exceeding $1/3$.
In addition, recently the trace anomaly \cite{Fujimoto:2022ohj}, $\Delta_{\rm tr} = 1/3 - P/\varepsilon $, attracts attention 
which contains information about the absolute value of $P$, in addition to the slope, at a given $\varepsilon$.
The vanishing of the trace anomaly implies the system being in the conformal limit.
The sound velocity peak and trace anomaly are the key quantities
to infer the relevant degrees of freedom, quarks or hadrons or else, 
and hence have crucial importance to characterize the state of matter. 

To test theoretical conceptions in dense QCD,
it is useful to study QCD-like theories such as two-color or isospin QCD \cite{Son:2000xc,Son:2000by,Splittorff:2000mm},
for which lattice simulations are doable without suffering from the sign problem.
For example, lattice simulations for two-color QCD \cite{Iida:2022hyy,Itou:2023pcl} have demonstrated the existence of the sound velocity peak,
and subsequently other simulations in isospin QCD also have confirmed it \cite{Brandt:2022hwy,Abbott:2023coj}. 

In a recent work \cite{Chiba:2023ftg}, we have analyzed a quark-meson model for isospin QCD
which should be suitable for the regime intermediate between hadronic and weakly coupled quark matter.
More precisely, a dense matter starts with the Bose-Einstein-Condensation (BEC) phase of pions
and smoothly changes into the Bardeen-Cooper-Schrieffer (BCS) phase with the pion condensates on top of the quark Fermi sea.
This can be regarded as the hadron-to-quark crossover in which the condensates (order parameters) have the same quantum number
but the qualitative features change.
We delineated the microphysics relevant to the sound velocity peak and trace anomaly.
Respecting the symmetry, 
the model results at low density are in good agreement with those from the Chiral effective field theory (ChEFT) \cite{Adhikari:2020qda,Lu:2019diy,GomezNicola:2022asf}
and lattice simulations \cite{Brandt:2022hwy,Abbott:2023coj}.
In the transient regime from hadronic to quark matter, 
we found that the sound velocity peak is developed  and remains larger than $1/3$ 
until the density reaches the pQCD domain, $\mu_I\sim 1$ GeV.
One of possible explanations for such slow reduction of $c_s^2$ is non-perturbative power corrections of $\sim +\lqcd^2 \mu_I^2$ to the pressure,
where the BCS gap of $\Delta \sim 200$-300 MeV appears in place of $\lqcd$ \cite{Chiba:2023ftg}.
The power corrections add negative contributions to the trace anomaly,
and hence the trace anomaly may be usable as a measure of non-perturbative effects in quark matter.
These conceptions are found to be qualitatively consistent with the lattice results of Ref.~\cite{Abbott:2023coj},
and, in some density interval, work even at quantitative level.

Based on our zero temperature analyses, 
we now proceed to analyses at finite temperature.
We add thermal quark contributions which couple to a Polyakov loop,
a measure of thermally generated color-field backgrounds.
Understanding the relationship between thermal effects and the phase structure 
should be important in the context of supernovae and binary NS mergers,
and also in heavy ion collisions at low energy, see Refs.~\cite{Sorensen:2023zkk,Lovato:2022vgq} for reviews 
and e.g., Refs.~\cite{OmanaKuttan:2022aml,Yao:2023yda} for up-to-date analyses.
How quark degrees of freedom affects these phenomena
has been studied most typically in models with first phase transitions from hadronic to quark matter \cite{Blacker:2023afl}.
In the phase transition region, a matter has $c_s^2=0$ and produces a latent heat.
In contrast, models with a crossover \cite{Masuda:2012kf,Masuda:2012ed,Masuda:2015kha,Kojo:2014rca,Kojo:2015fua,Baym:2017whm,Baym:2019iky,Kojo:2021wax}, 
which we are exploring, lead to the sound velocity peak
that leaves the astrophysical signatures differing from what we expect from purely nuclear or hybrid models with first order phase transitions \cite{Huang:2022mqp,Fujimoto:2022xhv,Kedia:2022nns,Harada:2023eyg}.
How stiffening, rather than softening, 
affects the above-mentioned phenomena is an important question
related to the structure of the QCD phase diagram.
We examine the sound velocity and trace anomaly along isentropic trajectories on which the $s/n_B$ is constant.

This paper is organized as follows.
In Sec.~\ref{sec:model} we briefly review the quark-meson model and the renormalization procedures.
The Polyakov loops are coupled.
In Sec.~\ref{sec:eos} we examine equations of state of isospin QCD, 
especially the behaviors of the sound velocity and trace anomaly
along isentropic trajectories.

\section{Model}
\label{sec:model}

\subsection{Renormalization}
\label{sec:renormalization}

The quark-meson model for isospin QCD, including the renormalization procedures, 
was discussed in our last paper \cite{Chiba:2023ftg} (which follows the original works in Refs.~\cite{Adhikari:2016eef,Adhikari:2018cea}),
so here we only briefly review the model and then discuss the modification for finite temperature.

In our treatment we include only the leading order of the $1/\Nc$ expansion.
Our thermodynamic potential is dominated by quarks and condensed mesons of $O(\Nc)$.
We neglect quantum and thermal meson fluctuations which give $O(1)$ contributions;
although the structure of mesonic excitations is affected at the leading order of $1/\Nc$ expansion,
such feedback does not show up in the thermodynamics of $O(\Nc)$.
Our setup and approximation are the same as Refs.~\cite{Adhikari:2016eef,Adhikari:2018cea}
which put focus on the structure of the phase diagram,
while in our study we examine the EOS in detail.

The Lagrangian of the two-flavor quark-meson model at finite isospin chemical potential $\mu_I $, 
which appears in place of up- and down-quark chemical potentials, $\mu_u = \mu_I$ and $\mu_d = - \mu_I$, 
is
\begin{align}
	\calL (\vec\phi)=&\, {1\over2}\left[(\partial_\mu \sigma)^2+(\partial_\mu \pi_{3} )^2\right] \notag\\
		&\hspace{-.2cm} + (\partial_\mu + 2 \rmi \mu_I\delta_\mu^0) \pi^+\left(\partial^\mu - 2 \rmi \mu_I\delta^\mu_0\right)\pi^- \notag\\
		&\hspace{-.2cm} -{m_{0}^2 \over 2} \vec\phi^2  -{\lambda \over 24} ( \vec\phi^2 )^2  + h \sigma  \notag\\
		&\hspace{-.2cm} + \overline{\psi} \left[ \rmi \Slash{\partial} + \mu_I \tau_3 \gamma^0 - g (\sigma + \rmi \gamma^5 \vec\tau \cdot \vec\pi) \right]\psi \,.		
\end{align}
We explain the values of parameters at the end of this section.
Here $\psi$ is  a quark field with up- and down-quark components $\psi = \big( u, d \big)^T$.
The  $\vec\phi = (\sigma,\vec\pi)$ are meson fields which correspond to the isospin $\bf 1$ and $\bf 3$ representations.
The $\tau_i$'s are the Pauli matrices in flavor space.
We note that our definition of $\mu_I$ gives $n_I = n_u - n_d$,
a factor 2 larger than the other conventional definition $n_I^{\rm conventional} = (n_u-n_d)/2$.

The model at $\mu_I=0$ and $m_\pi \neq 0$ has the flavor $U(2)_V$.
With $0 < \mu_I < m_\pi/2$, we are left with the $U(1)_{I_3} \otimes U(1)_B $ symmetry.
For $\mu_I \ge m_\pi/2$, the $U(1)_{I_3}$ symmetry is further broken down, 
yielding a single massless mode.
The dynamics of such a mode at low energy can be efficiently described by 
the effective field theory of the $U(1)_{I_3}$ breaking.

At tree level, we can already see that the $\mu_I$ favors the nonzero values of $\pi_{1,2}$, the pion condensates.
Without loss of generality one can choose $\la \pi_1 \ra  \neq 0$ and $\la \pi_2 \ra = 0$.
Including one-loop corrections from quarks changes the response of pion condensates to $\mu_I$ especially at high density,
since quarks become responsible for quantities such as isospin density which, at tree level, are solely carried by condensed pions.
The condensed mesons affect the quark energy through the Dirac mass and gap as 
\beq
M_q = g \la \sigma\ra\,,~~~~~~~ \Delta = g \la \pi_1 \ra \,,
\eeq
which, in the large $\Nc$ limit, is renormalization group (RG) invariant,
while $g$ and $\sigma$ are separately RG variant \cite{Chiba:2023ftg,Adhikari:2016eef,Adhikari:2018cea}.
The quark energies are
\beq
	E_u = E_{\overline{d}} = E(\mu_I)\,,~~~ E_d=E_{\overline{u}}=E(-\mu_I) \,,
\eeq
where 
\beq
	E(\mu_I) = \sqrt{\left(E_D-\mu_I\right)^2+\Delta^2} \,,
\eeq
with $E_D = \sqrt{ \vp^2 + M_q^2}$.

The effective potential can be concisely written in the $\MS$ renormalization scheme,
\beq
V_{\rm 1-loop}
&=& 
 \frac{\, m^2 \,}{\, 2 g^2 \,} \big( M_q^2 + \Delta^2 \big) 
 + {\, \lambda \, \over 24 g^4 } \big( M_q^2 + \Delta^2 \big)^2  
		-  \frac{\, h \,}{\, g \,} M_q  
 \notag \\
 &+& 
  \frac{\, 2N_c \,}{\,  (4\pi)^2 \,} \bigg( \frac{3}{2} -  \ln \frac{\, M_q^2 + \Delta^2 \,}{\, M_0^2 \,} \bigg) \big( M_q^2 + \Delta^2 \big)^2 
\notag \\
&-& 2 \mu_I^2 \bigg( \frac{\, 1 \,}{\, g^2 \,}  - \frac{\, 4 \Nc  \,}{\, (4\pi)^2 \,} \ln \frac{\, M_q^2 + \Delta^2 \,}{\, M_0^2 \,} \bigg) \Delta^2
 \notag \\
 &+& V_q^R (\mu_I, M_q, \Delta)  \,,
\eeq
where $M_0$ is the Dirac mass in vacuum and $(m, g, \lambda)$ are parameters evaluated in the $\MS$ scheme at a scale $M_0$.
The $V_q^R$ is the subtracted single particle energy,
$V_q^R = V_q - V_q^{(0)} - V_q^{(2)}$,
where
\beq
V_q &=& -N_c\int_{\vp} \big( E_u+E_d+E_{\overline{u}}+E_{\overline{d}} \big) \,.
\notag \\
V_q^{(0)} &=& - 4\Nc \int_{\vp} \sqrt{ E_D^2+\Delta^2} \,, 
\notag \\
V_q^{(2)} &=& - 2 \Nc \int_{\vp} \frac{\, \mu_I^2 \Delta^2 \,}{\, ( E_D^2+\Delta^2)^{3/2} \,} \,.
\eeq
The potential $V_q^R$ is well-saturated by $\mu_I^4$ terms and contains vanishing $\mu_I^2$ contributions.

The parameters in the $\MS$ effective potential can be related to 
the physical parameters in the on-shell scheme,
and we use the effective potential rewritten with the on-shell parameters,
see Ref.~\cite{Chiba:2023ftg} for the full expression.
We determine the parameters so as to reproduce 
$m_\pi = 140$ MeV, $m_\sigma = 600$ MeV, $f_\pi = 90$ MeV, and $M_q = 300$ MeV.
The details can be found in Ref.~\cite{Chiba:2023ftg}.

\subsection{Thermal contributions}
\label{sec:thermal_contributions}
 
As we have stated at the beginning of this section, 
we do not explicitly calculate the thermal mesonic contributions within our model.
The only contributions we take into account are thermal quark contributions. 
In principle mesonic excitations at low baryon density may be calculated within the hadron resonance gas picture \cite{Roessner:2007gha},
while excitations in the pion condensed phase,
where the isospin and parity symmetry are broken, 
require coupled channel calculations among $\pi$'s and $\sigma$.
In order to include the latter into the thermodynamics, 
we need not only the masses but also the full momentum dependence
and discussions for the dissociation scale of excitations.
Such extensive analyses are beyond the scope in this paper.
Instead, to gain some insights we give rough estimates of mesonic contributions
in the pion condensed phase in Sec.~\ref{sec:thermal_pions}.
 
The quark contributions are potentially $O(\Nc)$ but there are two important effects that suppress the thermal quark effects.
The first is the condensation effects with which quark excitations are gapped and acquire the Boltzmann factor of $\sim \rme^{-\Delta/T}$.
With a naive BCS estimate, $\Delta \sim T_c /0.57 \sim 300$ MeV where $T_c \sim 170$ MeV is the estimate from the lattice simulations \cite{Brandt:2022hwy}.

Even without the pion condensation,
an isolated quark excitation costs the large energy associated with a color electric flux (QCD string) emerging from the quark.
At high enough temperature, such a QCD string can be connected with color fields supplied by either thermally excited hadrons
or a quark-gluon-plasma.
Those colored background can be found when the entropic effects dominate over the energetic cost.
A measure of such colored background is a Polyakov loop;
it becomes finite at large temperature where QCD strings condense \cite{Polyakov:1978vu}.

Quark models coupled to Polyakov loops have given successful descriptions for the thermodynamics at zero chemical potential,
see Ref.~\cite{Fukushima:2017csk} for a review.
At low temperature, Polyakov loops offer extra suppression factors for thermal quark excitations, 
reflecting the extra energy cost associated with an isolated quark excitation.
This delays the chiral restoration and improves the agreement between quark model calculations 
and the lattice simulation results (at $\mu_B = 0$).

In the following we add the thermal contribution to the zero temperature expression as
\beq
V_{\rm 1-loop}^T 
\equiv V_{\rm 1-loop} + V_q^T + V_L \,,
\eeq
where $V_q^T$ is the thermal quark contribution and $V_L$ is the Polyakov loop potential.

The thermal quark part contains the coupling with the Polyakov loop and has the form
\beq
V_q^T = - 2 T \sum_{f= u,d, \bar{u}, \bar{d} } \int_\vp \ln w_f \,,
\eeq
with the Boltzman factors coupled to the Polyakov loop
\beq
&w_{u,d} = 1 +3\Phi \rme^{-\beta E_{u,d} } + 3 \bar{\Phi} \rme^{-2\beta E_{u,d} }+ \rme^{-3\beta E_{u,d} } 
\,, \notag \\
&w_{ \bar{u}, \bar{d} } = 1 +3 \bar{\Phi} \rme^{-\beta E_{ \bar{u}, \bar{d}} } + 3  \Phi  \rme^{-2\beta E_{\bar{u}, \bar{d}} }+ \rme^{-3\beta E_{\bar{u}, \bar{d}} } \,. 
\eeq
For $\Phi \rightarrow 0$ at $T\rightarrow 0$, only the last term survives while, at $\Phi \rightarrow 1$ at large $T$, 
the factor is summarized into $w \rightarrow (1+\rme^{-\beta E})^3$, recovering the usual form for deconfined quarks.

For the Polyakov loop potential, we use the parametrization motivated by the Haar measure \cite{Fukushima:2003fw,Roessner:2006xn}
\beq
&\hspace{-3cm}
T^{-4} V_{L} (\Phi,\Phi^*, T) 
= - a(T) \bar{\Phi} \Phi  /2 \notag \\
& + b(T) \ln\big[ 1 - 6 \bar{\Phi} \Phi + 4\big( \bar{\Phi}^{3} + \Phi^3 \big) -3\big(\bar{\Phi} \Phi \big)^2 \big] \,,
\eeq
where the $\Phi$'s are mean field values of the traced Polyakov loop,
\beq
\Phi  = \frac{1}{\, \Nc \,} \la \tr_{\rm c} L \ra \,,~~~~~ L=\calP \rme^{\rmi \int_0^\beta \rmd x_0 A_0 } \,.
\eeq
The parameters $a(T)$ and $b(T)$ is assumed to be
\beq
a(T) = a_0 + a_1/t + a_2/t^2 \,,~~~~b(T) = b_0/t^3 \,,
\eeq
with $t = T/T_0$.
The parameters are given as \cite{Roessner:2006xn}
\beq
\hspace{-0.5cm}
a_0 = 3.51\,,~a_1= -2.47\,,~ a_2=15.2\,,~ b_0 = -1.75\,.
\eeq
The parameter $T_0$ is chosen to be \cite{Schaefer:2007pw}
\beq
T_0 (\Nf) = T_\tau \rme^{-1/\alpha_0 \gamma}\,,
~~~~~
\gamma = \frac{\, 11\Nc - 2\Nf \,}{6\pi} \,,
\eeq
with $\alpha_0 = 0.304$ and $T_\tau = 1.77$ GeV.
The present parameterization leads to $T_0(\Nf=0) = 270$ MeV for the pure glue theory,
$T_0(\Nf=2) = 208$ MeV, and $T_0(\Nf=3) = 178$ MeV for finite quark flavors.
With these parameters, the critical temperature for the chiral restoration in $\Nf=2$ theories 
appears to be $T_c \sim 180$ MeV \cite{Schaefer:2007pw}.

In our analyses at finite density, we use the same parameter sets as the vacuum case unless otherwise stated.
In the literatures there are versions with the $\mu_I$ dependence of $\gamma$,
e.g., $\gamma \rightarrow \gamma - 16\Nf \mu_I^2/\pi T_\tau^2$, originally introduced in Ref.~\cite{Schaefer:2007pw} with considerations based on the hard-dense loop. 
In the normal phase, this introduces $\sim 10\%$ reduction of the critical temperature.
Meanwhile the situation may change in the pion condensed phase;
the back reaction from quarks to the gluon sector is expected to be suppressed due to the color-singlet BCS gap.
This conjecture has been tested for the gluon propagators in two-color QCD \cite{Kojo:2014vja,Suenaga:2019jjv,Kojo:2021knn,Contant:2019lwf}.
There it is found crucial to use quark propagators with the BCS gap to explain the lattice data \cite{Boz:2018crd,Bornyakov:2020kyz}.
Combining gluon propagators weakly depending on $\mu_I$ with the computation of the effective potential for Polyakov loops \cite{Fukushima:2012qa},
we expect that the $\mu_I$ dependence to be included in the parameter $\gamma$ should be modest.
In what follows, we have checked that 
using the $\mu_I$-dependent $T_0$ introduces $\sim 10\%$ enhancement in the critical temperature
for the domain of our interest (see also Ref.~\cite{Adhikari:2018cea}),
and we present only the simplest version of the model with $T_0$ fixed as the vacuum case.


%

\section{Equations of state}
\label{sec:eos}

We first examine the gross features of the phase digram
in the $\mu_I$-$T$ plane
and look into the equations of state.

\subsection{Phase structure}
\label{sec:phase_structure}

\begin{figure}[tbp]
\vspace{-0.cm}
\centering
\includegraphics[width=1.0\linewidth]{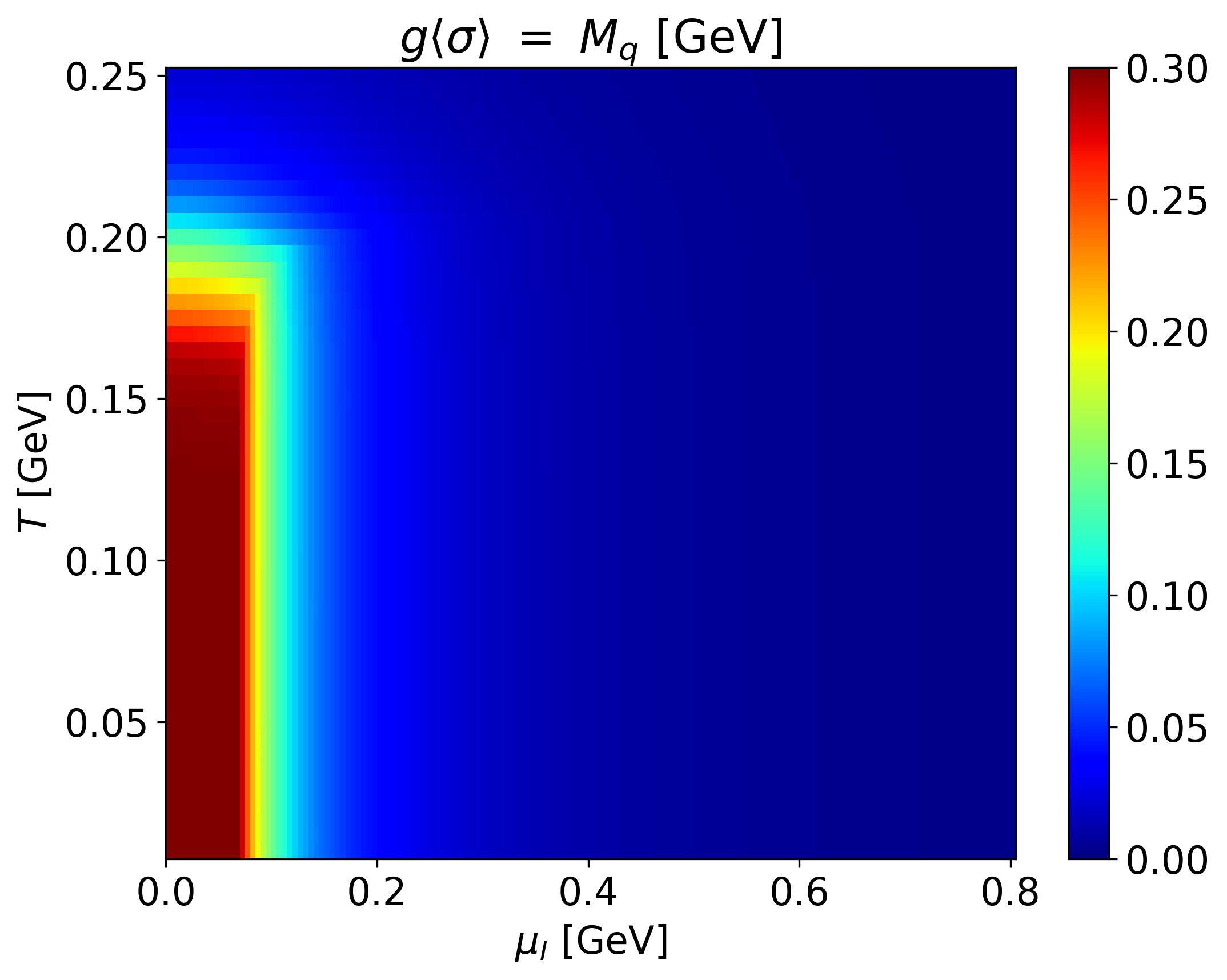}
\includegraphics[width=1.0\linewidth]{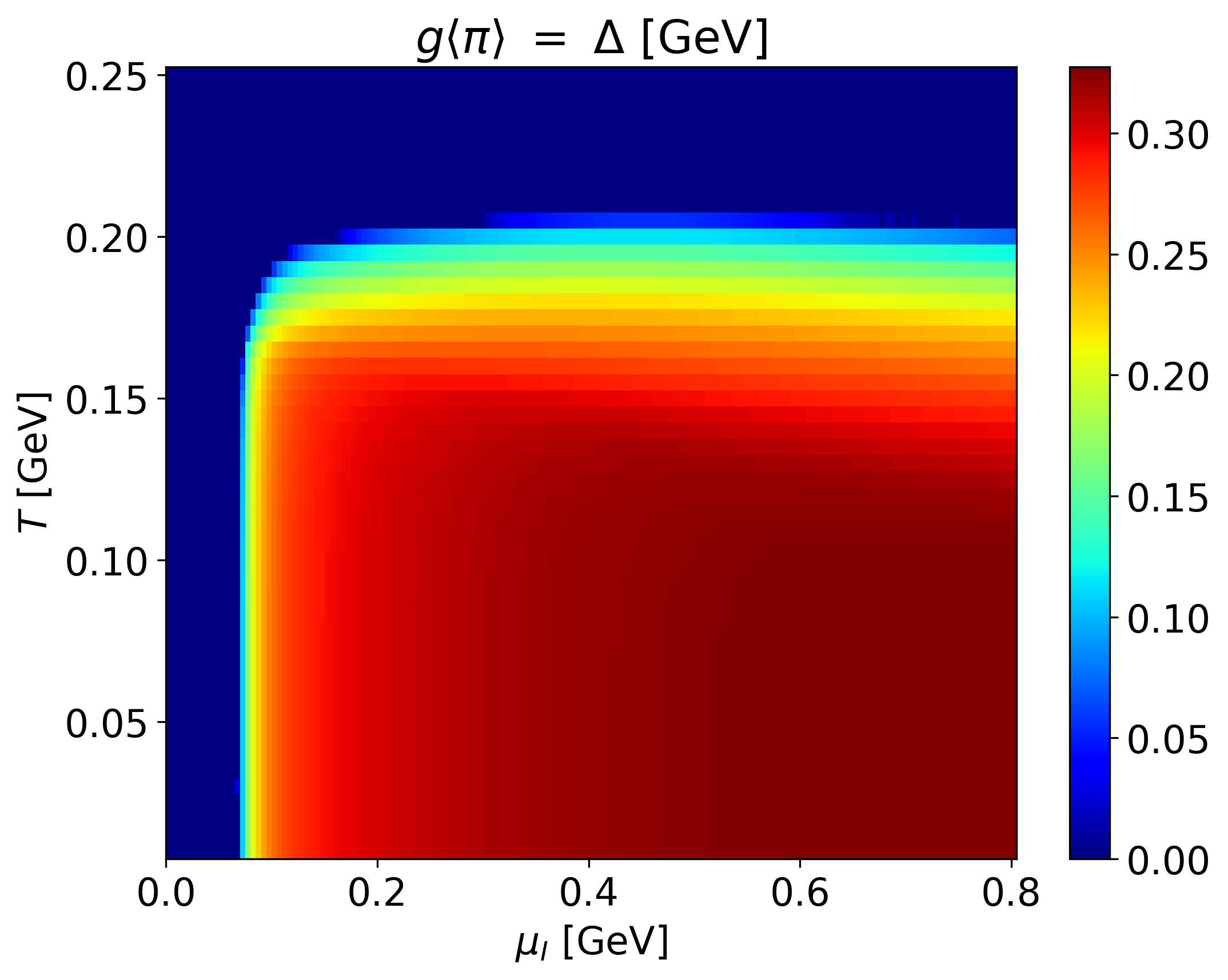}
\includegraphics[width=1.0\linewidth]{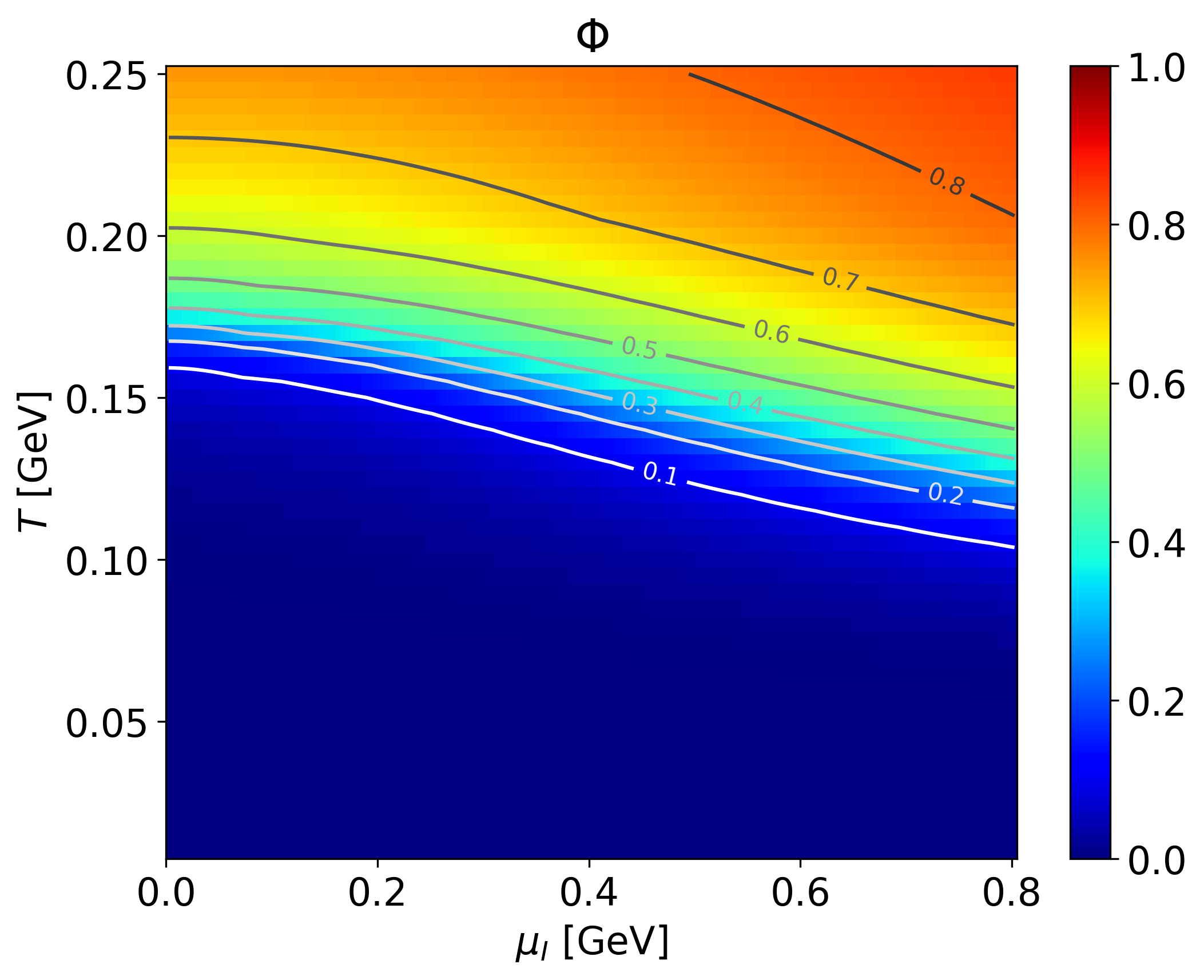}
\caption{ The phase structure:
	{\bf Top}: dynamical quark mass $M_q = g\langle \sigma \rangle$;
	{\bf Middle}: quark mass gap $\Delta = g \langle \pi_1 \rangle$;
	{\bf Bottom}: Polyakov loop.
		}
\label{fig:phase_structure}
\end{figure}

First we take a look at the chiral and pion condensates and Polyakov loop at given $(\mu_I, T)$.
Up to $\mu_I \simeq m_\pi/2$, 
the phase structure is similar to what has been calculated for the hadronic phase.
As shown in Fig.~\ref{fig:phase_structure}, 
at $T=0$ and beyond $\mu_I = m_\pi/2$, the charged pions condense while the chiral condensates begin to dissociate
but approach zero only asymptotically.
The phase transition line at $\mu_I \sim m_\pi/2$ is almost constant up to $T\sim 150$ MeV.
In the mean field calculations, the transition lines separating the pion condensed phase and the others
are the second order phase transition lines.

Without the Polyakov loops, the critical temperature for the pion condensate, $T_c^{\pi}$, is $\simeq 170$ MeV.
Including the Polyakov loop enhances $T_c^{\pi}$ to $\simeq 200$ MeV, about 30 MeV larger than the lattice simulation results.
This overestimate is in part duel to our neglect of thermal contributions from in-medium hadrons.
In Sec.~\ref{sec:thermal_pions} we try to give rough estimates for hadronic excitations.
Another problem in our current modeling is that we are not including the medium modifications of the Polyakov loop potential
which we have mentioned in the previous section.
In what follows we proceed with the current setup 
while keeping in mind that thermal corrections to EOS would be underestimated.

At a fixed temperature, the Polyakov loop values decrease as $\mu_I$ increases.
Within our one-loop computations in the large $\Nc$ treatment,
such reduction at larger $\mu_I$ is due to the phase space enhancement of $\sim 4\pi p_F^2$ ($p_F \sim n_I^{1/3}$: the quark Fermi momentum) for thermal quarks,
provided that the excitation gaps remain similar from low to high $\mu_I$.
Actually in our model such a constant gap is realized at low temperature.

\subsection{Equations of state}
\label{sec:eos}

\begin{figure}[tbp]
\vspace{-0.cm}
\centering
\includegraphics[width=1.0\linewidth]{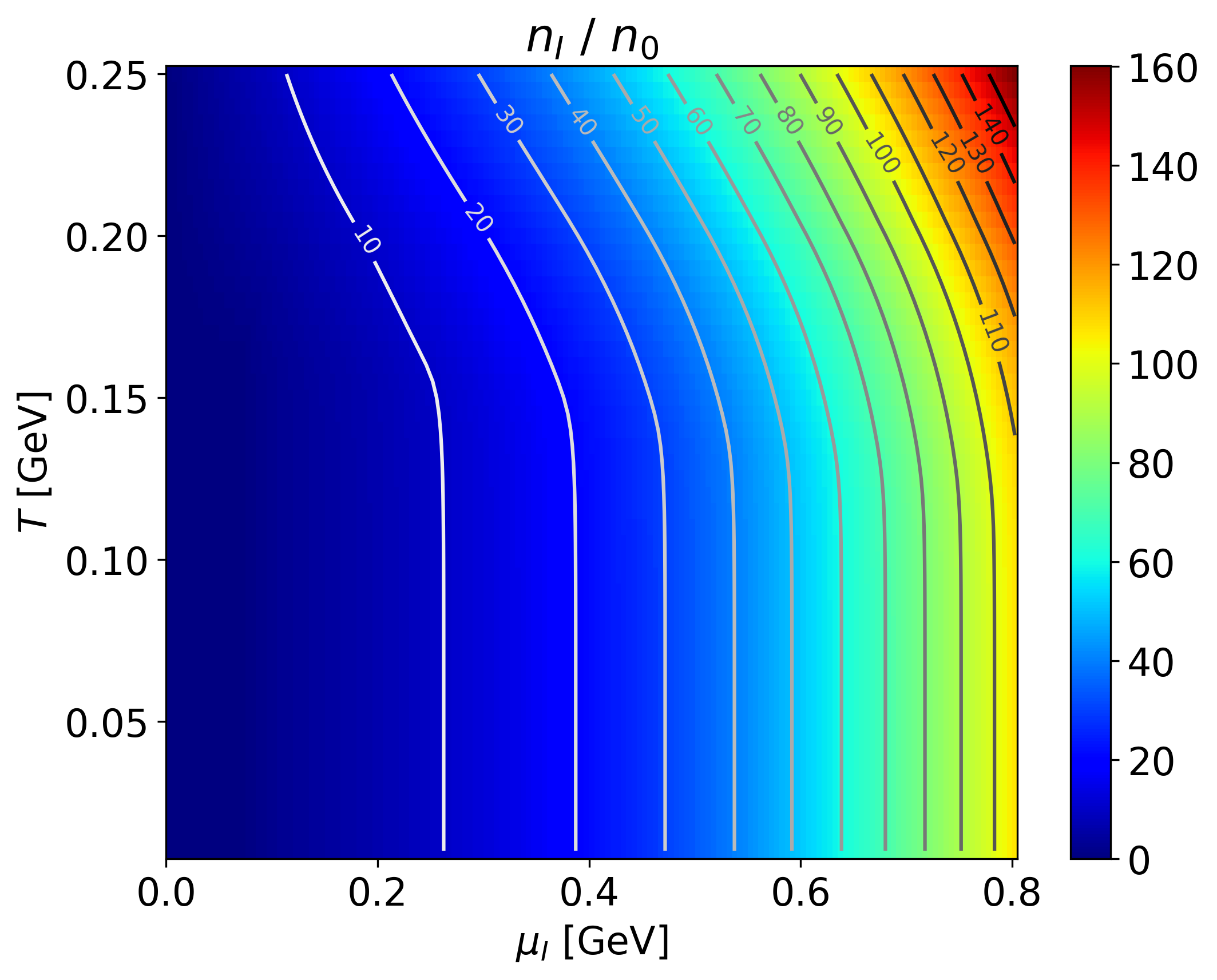}
\includegraphics[width=1.0\linewidth]{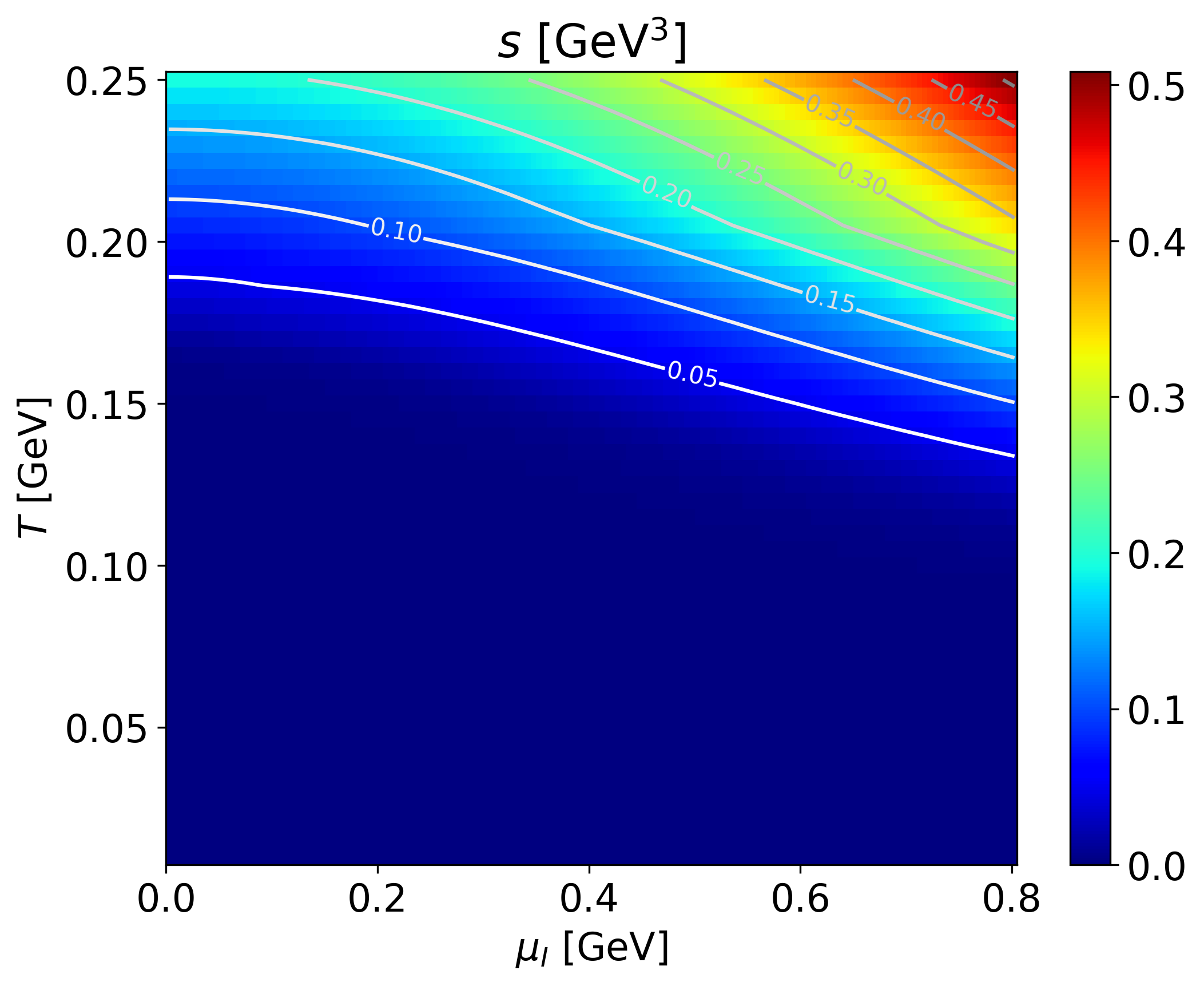}
\caption{ The $n_I/n_0$ (top) and $s$ (bottom) in the $\mu_I$-$T$ plane.
}
\label{fig:table_n-s}
\end{figure}

\begin{figure}[tbp]
\vspace{-0.cm}
\centering
\includegraphics[width=1.0\linewidth]{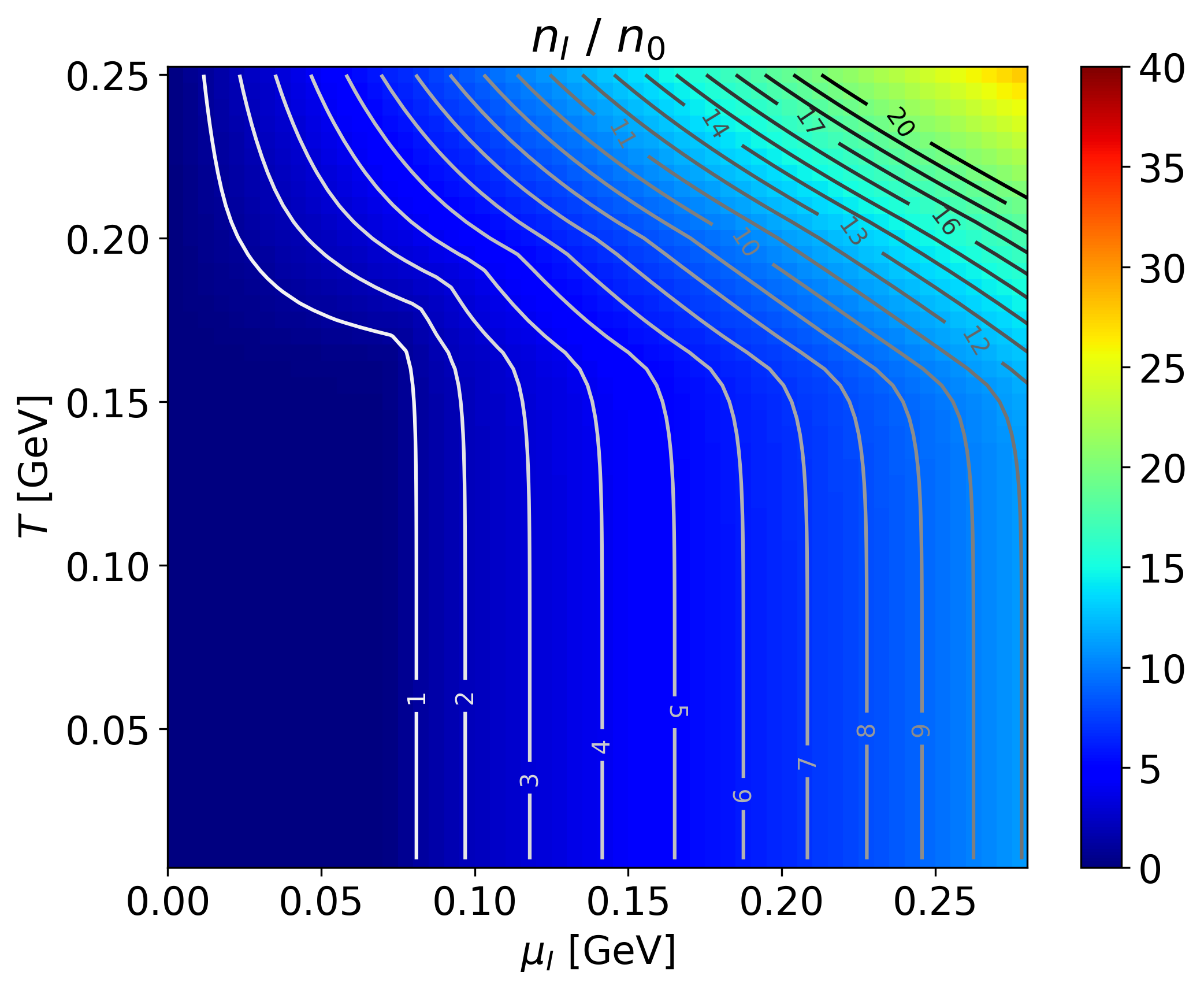}
\caption{ The $n_I/n_0$ with more focus on the low density region 
at $\mu_I \le 2m_\pi$.
}
\label{fig:table_n_tight}
\end{figure}

We study EOS mainly for the isentropic trajectories, i.e., EOS at fixed $s/n_I$.
Before doing so, we first show $n_I$ and $s$ as functions of $\mu_I$ for several $T$ as a preparation.

Shown in Fig.~\ref{fig:table_n-s} are the isospin density $n_I/n_0$ and entropy density $s$.
For $n_I$, in Fig.~\ref{fig:table_n_tight} we also show the same plot but with more focus on the low density domain at $\mu_I \le 2m_\pi$.
The experimental determination based on the $\pi e$ scattering and the $e^+ e^- \rightarrow \pi^+ \pi^-$ process \cite{Ananthanarayan:2017efc} 
yields the estimate
$r_\pi^V = \sqrt{ \la r^2 \ra_V } ~\simeq~ 0.66\, {\rm fm}$ \cite{PhysRevD.98.030001}
which has been reproduced by lattice calculations \cite{Koponen:2015tkr,Wang:2020nbf}.	
The isospin density where pions overlap is estimated as
(reminder: our definition of $n_I$ is $n_u - n_d$, so $n_I/2$ should be equated with the inverse volume of pions with isospin 1)
\beq
n_I^{\rm overlap} = 2 \times \big( 4\pi r_\pi^3/3)^{-1}  \simeq 10.4 n_0 \,, 
\eeq
which corresponds to $\mu_I \simeq 0.26\, {\rm GeV}$.

At finite temperature and low density, 
the overlap of thermally excited hadrons is roughly characterized by the entropy of 2-3 $ {\rm fm}^{-3}$,
supposing that a single meson or baryon including two or three quarks is available within a volume $\sim 1\, {\rm fm}^{3}$.
%
That is, at $s \sim$ 2-3 $ {\rm fm}^{-3}$ thermally excited hadrons should begin to overlap each other.
Lattice simulations at $\mu_I =0$ show that 
the entropy density around the critical temperature $T_c\simeq 155$ MeV is $s \sim 0.02\, {\rm GeV}^3 \simeq 2.6\, {\rm fm}^{-3}$
which seems consistent with the naive estimate.

Next we consider the entropy at high density. 
As a guide we first consider the entropy in the Stefan-Boltzmann limit ($\Nc=3, \Nf=2$),
\beq
s_{q}^{ {\rm SB} }
= \frac{\,\Nc \Nf \,}{\, 6 \,}\bigg(
\mu_I^2 T  +  \frac{\, 7\pi^2 T^3 \,}{15} 
 \bigg) \,,
\eeq
where quarks are treated as gapless.
The $\mu_I^2$ factor in the first term reflects the phase space around the Fermi surface.
For $\mu_I = 1$ GeV $\simeq 7 m_\pi$, the entropy of $s \simeq$ 2-3 ${\rm fm}^{-3}$
would be reached at $T \sim 20$ MeV with gapless quarks;
a large entropy could be reached at very low temperature.

The situation is very different in the pion condensed phase.
Here thermal quark excitations are suppressed by the gaps.
The entropy at low $T$ parametrically scales as $s_q \sim \mu_I^2 T \rme^{-\Delta/T}$.
Hence EOS at low temperature, up to $T\sim 150$ MeV, and $\mu_I \lesssim 7m_\pi$ remain almost the same as the $T=0$ case.

\begin{figure}[tbp]
\vspace{-0.cm}
\centering
\includegraphics[width=1.0\linewidth]{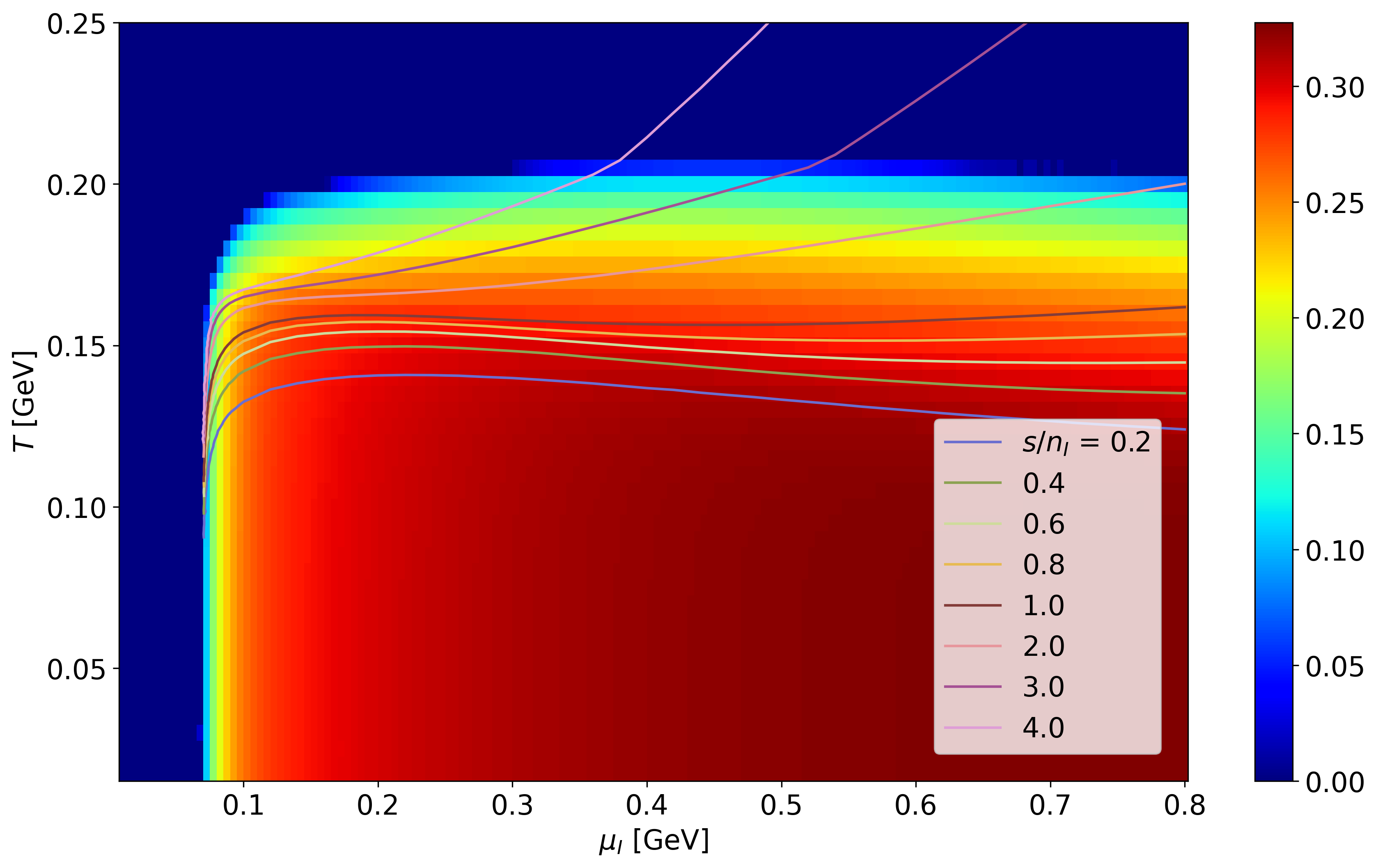}
\includegraphics[width=1.0\linewidth]{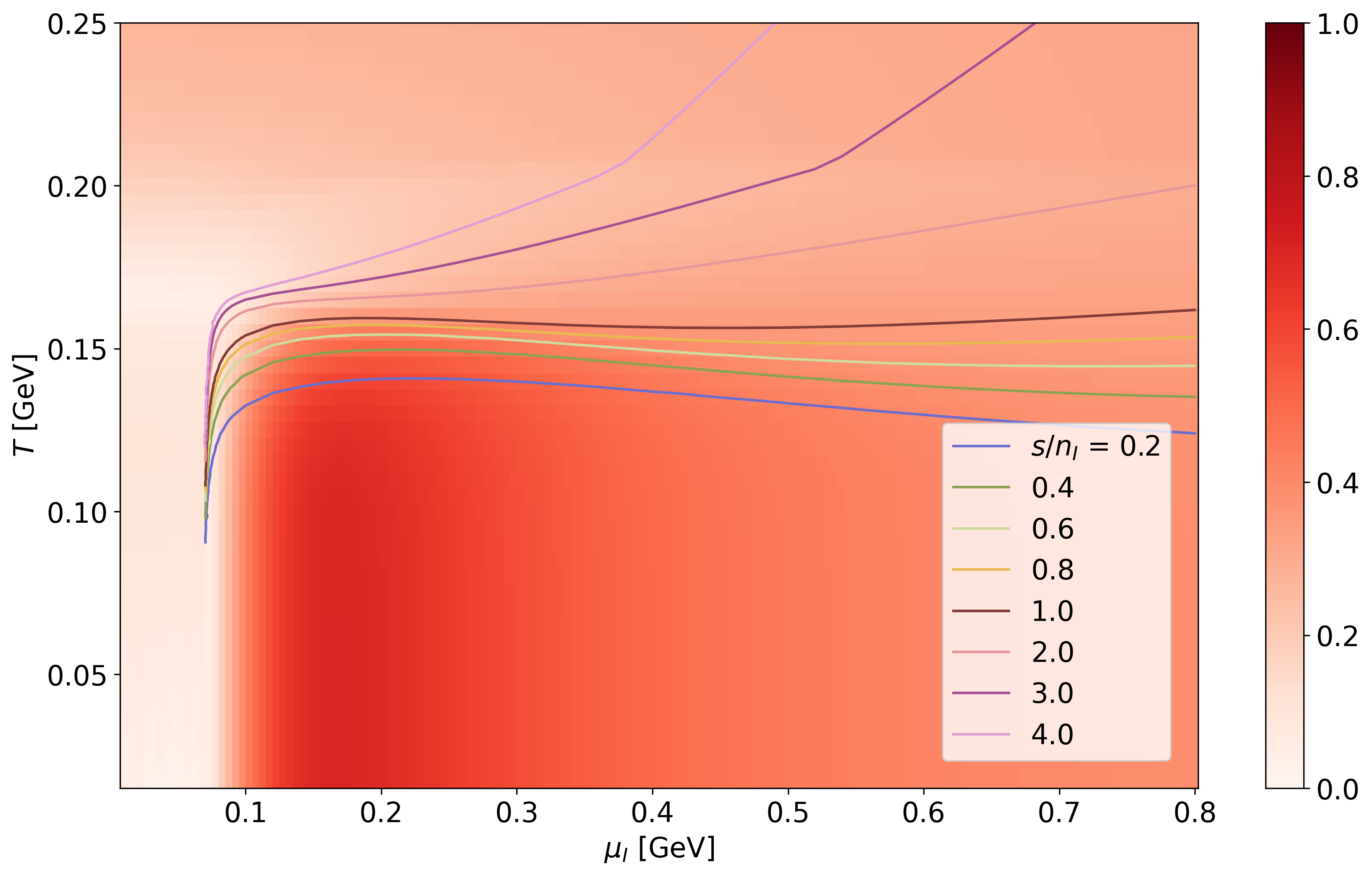}
\caption{ The contour for $\Delta$ (top) and $c_{s/n_I}^2$ (bottom) and isentropic (fixed $s/n_I$) trajectories in the $\mu_I$-$T$ plane.
The thickness of the cells represents the size of $\Delta$ and $c_s^2$.
The kinks in the isentropic trajectories around $T\simeq 200$ MeV
are due to the dissociation of pion condensates (the second order phase transition).
		}
\label{fig:isentropic}
\end{figure}

Now we examine the isentropic trajectories in the $\mu_I$-$T$ plane, as shown for $s/n_I =$ 0.2 to 4.0, 
together with the squared sound velocity along the trajectory, $c_{s/n_I}^2 = (\partial P/\partial \varepsilon)|_{s/n_I\, {\rm fixed} }$, 
in Fig. \ref{fig:isentropic}.
These relatively low $s/n_I$ are achieved only when $s$ is very low or $n_I$ is large.

First we examine the low $\mu_I$ region with $\mu_I \sim m_\pi/2$ 
where the trajectories go from low to high $T$ without much change in $\mu_I$.
This domain is slightly above the threshold of the pion condensation beyond which $n_I$ grows as $\sim f_\pi^2 \mu_I$,
as can be derived from the ChEFT.
Meanwhile the entropy calculated based on thermal quark excitations (without thermal pions)
is controlled by a factor $\sim \rme^{-\Delta(\mu_I, T)/T}$
where the gap near $\mu_I = m_\pi/2$ is sensitive to $\mu_I$.
A parametric expression is
\beq
s/n_I \simeq ({\rm phase~space}) \times  \rme^{ - \Delta(\mu_I, T)/T } \,,
\eeq
that is, the entropy becomes sizable only when $T$ is as large as $\Delta$.
More explicitly, the temperature scales as
\beq
 T (\mu_I )|_{ s/n_I }
\simeq  \frac{\, \Delta(\mu_I, T) \,}{\, \ln \big[\, ({\rm phase~space}) \times \big( n_I/s \big) |_{\rm fixed} \, \big] \,} \,
\eeq
where the denominator is logarithmic and its variation with respect to $ \mu_I, T$, and $ (s/n_I)_{\rm fixed}$  is modest.
Then, along the isentropic trajectories 
the $\mu_I$-dependence of $T$ is primarily determined by the behavior of the gap.

Next we consider the pion condensed domain at large $\mu_I$.
Until thermal corrections to $\Delta$ become substantial, 
$\Delta$ is almost constant for changes in $\mu_I$.
Applying the same discussion as the above, 
we conclude that $T $ in the isentropic trajectories shows the weak $\mu_I$-dependence,
as seen in e.g., $s/n_I \sim 0.8$ trajectory.
Meanwhile, when $\Delta \ll T$, the Boltzmann factor is approximately $\rme^{-\Delta /T} \sim 1$.
At large $\mu_I$,  $s \sim \mu_I^2 T$ and $n_I \sim \mu_I^3$ so that 
along the isentropic trajectories $T \propto \mu_I$.
This behavior is clearly seen in Fig.~\ref{fig:isentropic}.

\subsection{Sound velocity}
\label{sec:sound_velocity}

\begin{figure}[tbp]
\vspace{-0.cm}
\centering
\includegraphics[width=1.0\linewidth]{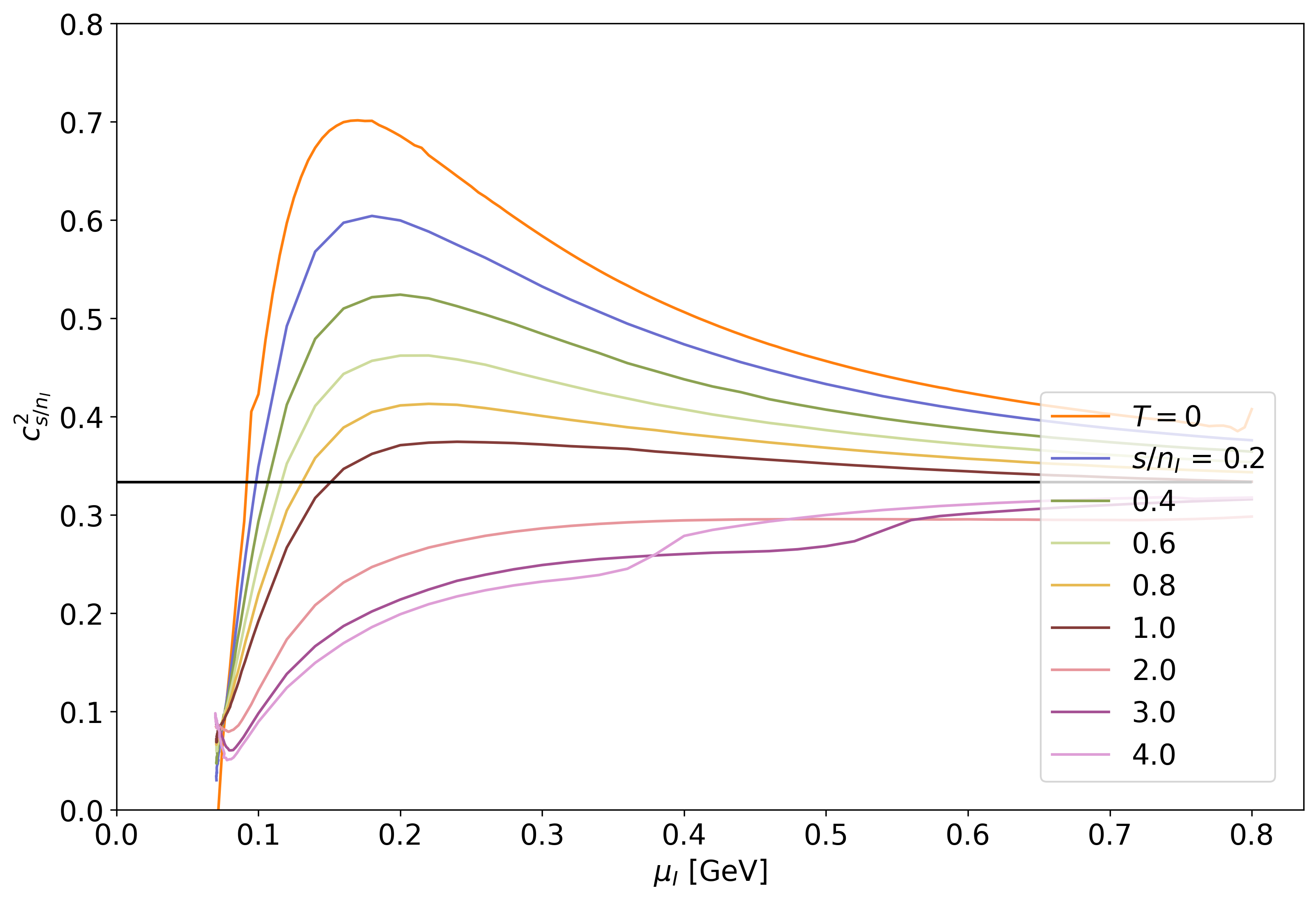}
\caption{ The squared isentropic sound velocity $c_{s/n_I}^2$ on isentropic trajectories.
At low $s/n_I$, a $c_s^2$ makes a peak as in the $T=0$ result.
As $s/n_I$ increases, the height of the peak decreases,
and the entire curve goes below the conformal limit $c_s^2=1/3$
for $s/n_I \simeq 1.5$.
		}
\label{fig:cs2_snap}
\end{figure}

Shown in Fig.~\ref{fig:cs2_snap} are the squared sound velocity $c_s^2$ along various isentropic trajectories,
from $s/n_I=0.2$ to $3.5$.
For a small $s/n_I$, the sound velocity in the BEC-BCS crossover region 
has a peak with the height greater than the conformal value $c_{s/n_I}^2=1/3$.
Increasing $s/n_I$, the trajectories leave the pion condensed domain,
and the peak structure disappears.

For a normal quark gas, the $c_{s/n_I}^2$ does not exceed the conformal limit $1/3$ but asymptotically approaches it from below.
The same is true for a pure pion gas.
Thus, the $c_{s/n_I}^2$ exceeding $1/3$ should be related to condensed pions \cite{Hippert:2021gfs}.\footnote{Relations between the sound velocity peak and the chiral-partner contributions on top of the condensates in two-color QCD are discussed in Ref.~\cite{Kawaguchi:2024iaw}.}
Leaving from the pion condensed phase, 
the behaviors of $c_{s/n_I}^2$ approach the quark gas behavior
(reminder: we have not included thermal pions yet).

Thus, what is nontrivial is the $c_{s/n_I}^2$ in the pion condensed phase.
The sound velocity at $T=0$ is found to have a peak around $ \mu_I \sim m_\pi $.
Parametrically, the existence of the peak can be inferred by
combining the ChEFT at LO and the relativistic limit.
Near $\mu_I \sim 0.5 m_\pi$, one can use the EOS derived from the ChEFT at leading oder,
\beq
P_{ {\rm EFT} }^{\rm LO} 
= 2 f_\pi^2 \mu_I^2 \bigg( 1 - \frac{\, m_\pi^2 \,}{\, 4\mu_I^2 \,} \bigg)^2 \,,
\eeq
with which the sound velocity behaves as
%
\beq
c_{s}^2 \big|_{ {\rm EFT} }  = \frac{\,  \big( 2\mu_I \big)^4 - m_\pi^4 \,}{\,  \big( 2\mu_I \big)^4 + 3 m_\pi^4 \,} \,.
\eeq
The $c_s^2$ reaches $1/3$ at $\mu_I = m_\pi/( 2 \times 3^{1/4} ) \simeq 0.7 m_\pi$ 
and asymptotically approaches $1$ toward very large $\mu_I$.
But the ChEFT should lose the validity at $\mu_I \gg m_\pi$
and the EOS should be replaced with the quark EOS.
A useful measure to examine the impact of the quark substructure is
the occupation probability of quark states
\cite{Chiba:2023ftg,Kojo:2021ugu,Kojo:2021hqh,Fujimoto:2023ioi};
the existence of the upper bound imposes a constraint
which is invisible from purely hadronic models
and yields statistical repulsion among hadrons.

At asymptotically large $\mu_I$, the quark kinetic energy should dominate over interactions 
and then $c_s^2 = 1/3$ follows.
Reducing $\mu_I$ toward $\mu_I \sim m_\pi$,
the pQCD corrections predict $c_s^2$ is reduced from $1/3$.
To describe growing $c_s^2$ toward low density, we need other elements.
One possible candidate is the non-perturbative power corrections producing $+\lqcd^2 \mu_I^2$ terms with $\lqcd \sim$ 200-300 MeV.
The BCS gap $\Delta$ can appear in the place of $\lqcd$.
Now parameterizing ($a \sim O(1) > 0 $)
\beq
P_{\rm BCS} (\mu_I) = C \big( \mu_I^4 + a \Delta^2 \mu_I^2 \big)\,,
\eeq
we obtain 
\beq
c_{s}^2 \big|_{\rm BCS} 
= \frac{\,  2 \mu_I^2 + a \Delta^2 \,}{\, 6 \mu_I^2 + a \Delta^2 \,} \,.
\eeq
At $\mu_I \sim \Delta$, the $c_s^2$ can be substantially larger than $1/3$.

The quark-meson model covers both low and high density characteristics.
At low density condensed pions dominate the EOS,
while at high density condensed pions become responsible for the BCS gaps for quarks.

Finally, we notice that enhancement of $c_{s/n_I}^2$ at large $s/n_I \gtrsim 2.0$ and large $\mu_I \gtrsim 300$ MeV
which is related to melting of pion condensates.
As can be seen in Fig.~\ref{fig:isentropic}, 
these trajectories start with $T \gtrsim 150$ MeV at low density and have larger temperatures for larger $\mu_I$,
eventually crossing the critical line for pion condensed phase.
For these trajectories with $T \gtrsim 150$ MeV,
there are already substantial thermal quark excitations and sizable Polyakov loop contributions.
It is worth mentioning that, with a larger $\Delta$, thermal excitations are closer to the 
non-relativistic regime
\beq
\sqrt{ (E_q - \mu_I)^2 + \Delta^2 }
\simeq \Delta + \frac{\, v_F^2 \vp^2 \,}{\, 2\Delta \,}
\eeq
where we use an approximation $|E_q - \mu_I | \simeq v_F |\vp|$
for excitations near the Fermi surface.
Thus, while the presence of $\Delta$ enhances $c_s^2$ in the zero temperature EOS,
it acts oppositely for thermal corrections.
When $\Delta$ vanishes, thermal quarks become more relativistic and increases $c_s^2$.
\subsection{Trace Anomaly }
\label{sec:trace_anomaly}

A useful quantity to examine the conformality of the system
is the trace anomaly divided by the energy density \cite{Fujimoto:2022ohj}
\beq
\Delta_{\rm tr} = \frac{\, \la T^{\mu}_{\mu} \ra \,}{\, 3\varepsilon \,} = \frac{1}{\, 3 \,} - \frac{\, P \,}{\, \varepsilon \,} \,.
\eeq
The $\Delta_{\rm tr}$ is vanishing in the conformal limit.
We are mainly interested in the sign.

This quantity in isospin QCD has been studied at $T=0$ in Ref.~\cite{Chiba:2023ftg}.
Within pQCD, the perturbative corrections favor positive trace anomaly.
Then, we examine parametrized non-perturbative corrections.
The bag constant favors positive trace anomaly.
On the other hand, we found that attractive correlations near the Fermi surface yield 
positive power corrections of $\Delta^2 \mu_I^2$ that favor the negative trace anomaly.
Our prediction of the negative trace anomaly to high density turns out to be
consistent with the lattice calculations in Ref.~\cite{Abbott:2023coj} 
in which the trace anomaly is negative
up to $\mu_I \sim 7.5 m_\pi \simeq 1.28$ GeV, 
to very high density where pQCD has been supposed to be valid. 
This deviation from pQCD suggests that
non-perturbative effects survive from the hadronic to weakly coupled quark matter domain.

In this work we extend the analyses to finite temperature.
For $\mu_I\simeq 0$ and finite $T$,
it has been known that the trace anomaly is positive
and has a peak along $T$-axis around where a hadron resonance gas transforms into a QGP.
We examine when and how the negative trace anomaly found at $T=0$ changes the trend toward high temperature.

First we consider a normal phase at $\mu_I \lesssim m_\pi/2$.
In the non-relativistic regime with $\varepsilon \gg P$,
the trace anomaly is approximately $\la T^{\mu}_{\mu} \ra \simeq \varepsilon - 3P \simeq \varepsilon$ and hence is positive.
At low $\mu_I$ and low $T$, thermal quark excitations with the mass $m$ are highly suppressed by $\sim \rme^{-m/T}$.
Increasing $T$, each excitation becomes more relativistic so that the pressure becomes larger
and $\Delta_{\rm tr}$ approaches the conformal limit from above, $\Delta_{\rm tr} \propto \varepsilon - 3 P \rightarrow 0^+$.

For $\mu_I \gtrsim m_\pi/2$ and low $T$,
pion condensates are formed.
Near the onset, $\varepsilon \sim m_\pi n_I \gg P$ so that the trace anomaly is positive.
But with slight increase of $\mu_I$, 
the pressure develops rapidly exceeding the conformal value $\varepsilon/3$, 
and then stays beyond the conformal limit for a wide interval.
Increasing $T$, the trace anomaly does not change much within the pion condensates,
but begins to increase as the condensates melt.
In the range of $T = 0.15$-$0.20$ GeV, the trace anomaly becomes positive everywhere.

\begin{figure}[tbp]
\vspace{-0.cm}
\centering
\includegraphics[width=1.0\linewidth]{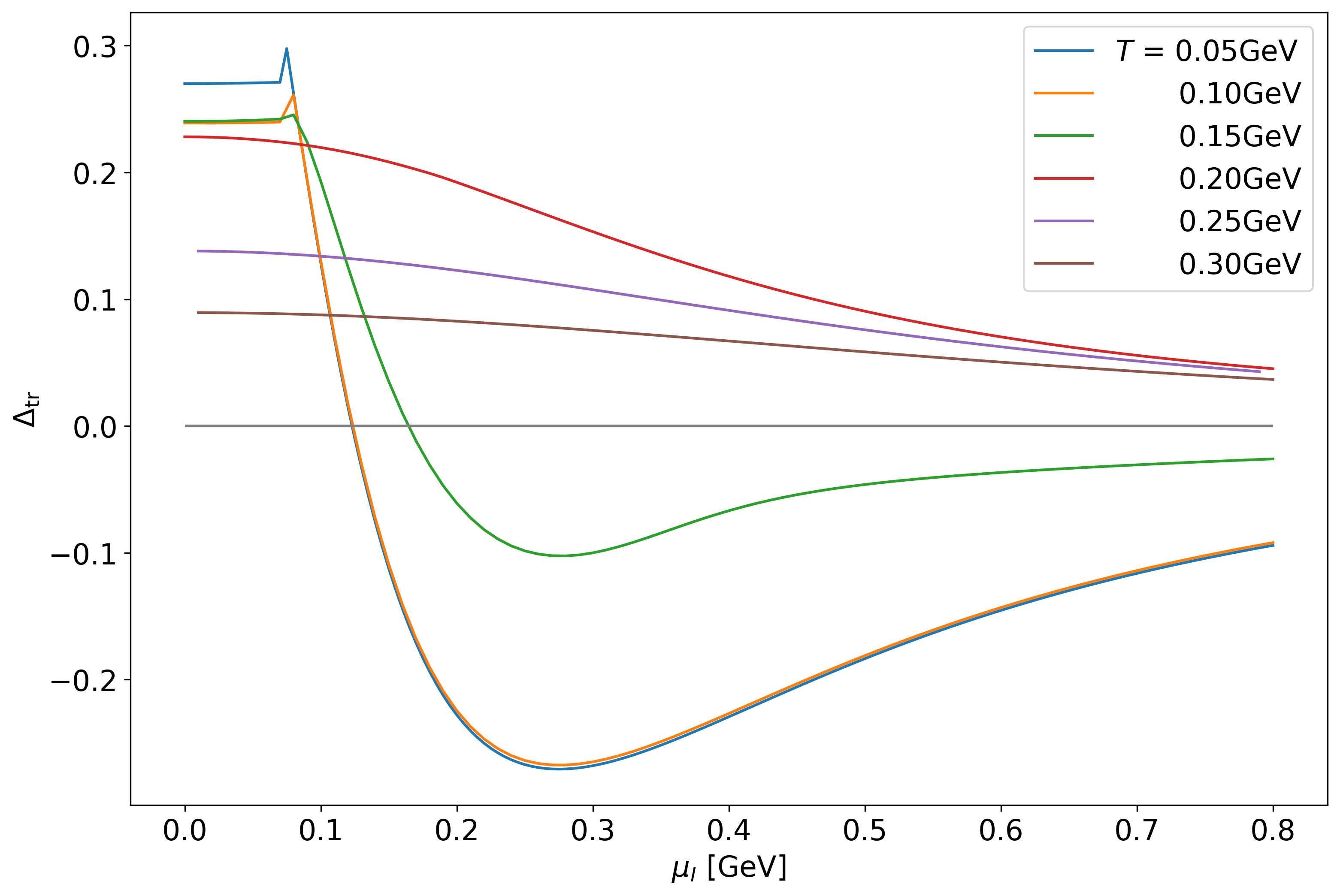}
\caption{ The trace anomaly $\Delta_{\rm tr}$ as a function of $\mu_I$ at various temperatures. 
In the non-relativistic limit, $\Delta_{\rm tr} \simeq 1/3$.
		}
\label{fig:tr}
\end{figure}

\subsection{Thermal pions: rough estimates}
\label{sec:thermal_pions}

%
\begin{figure}[tbp]
\vspace{-0.cm}
\centering
\includegraphics[width=1.0\linewidth]{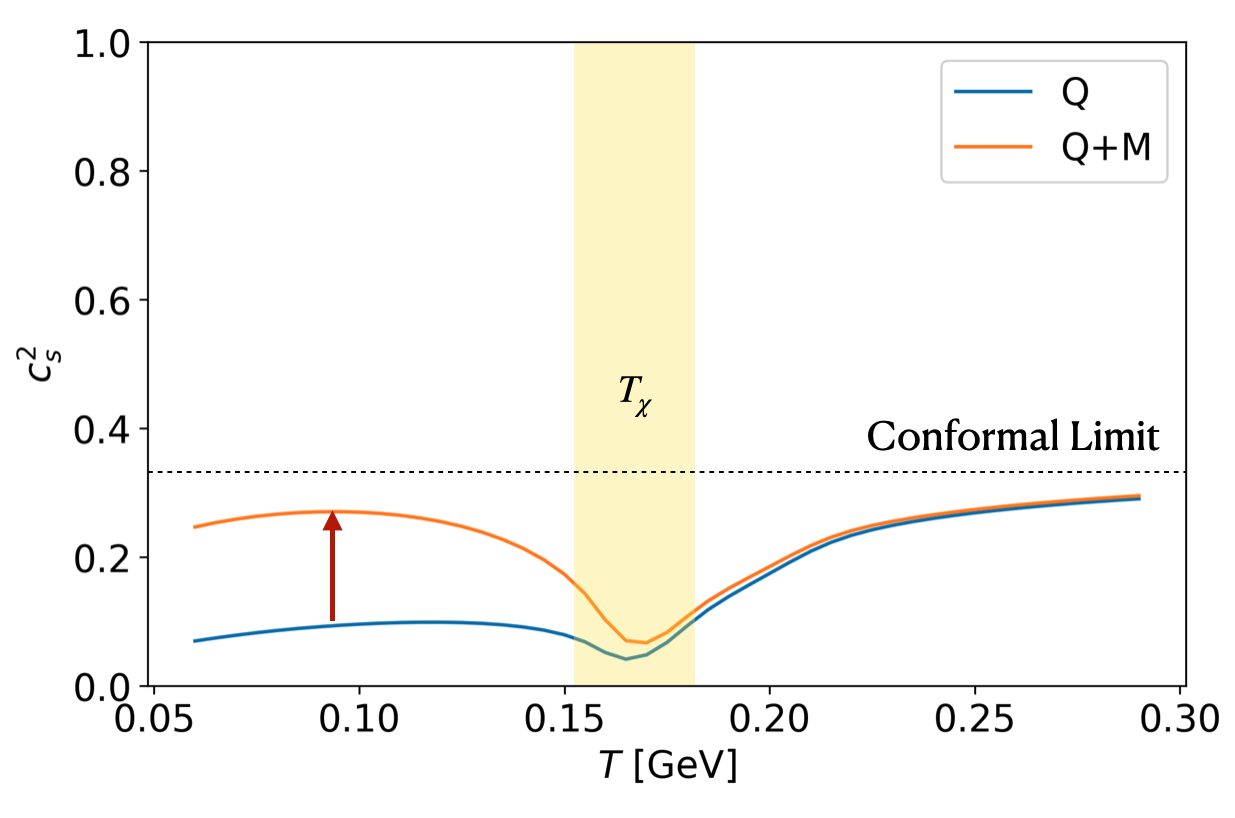}
\caption{ The squared sound velocity $c_s^2$ at $\mu_I=0$ and various temperatures.
	At low temperature, the EOS including only thermal quarks underestimate $c_s^2$ and including thermal pions significantly enhance it.	
	The window for the chiral restoration is shown as a guidance. 
	The window for deconfinement, characterized by the Polyakov loop, appears to be in the same range as $T_\chi$ and hence is not shown explicitly.
}
\label{fig:cs_sn_10.png}
\end{figure}
%

Our quark-meson model within the one-loop approximation misses thermal hadronic contributions.
At low density, this causes serious errors in estimating the thermodynamic quantities.
In fact, it has been known that thermal pions are dominant at low temperature.
Indeed, typical quark models lead to too small sound velocity at low temperature
since thermal quarks are massive and stay within the non-relativistic regime (Fig.~\ref{fig:cs_sn_10.png}).
Pions, which are lighter than quarks, reach the relativistic regime earlier
and hence increase $c_s^2$ to the value close to the conformal limit.
Approaching the chiral restoration temperature $T_\chi \simeq 160$ MeV,
the massive quarks dominate over pions in the EOS and the system goes back to the non-relativistic regime,
creating a dip in $c_s^2$ before reaching the relativistic or conformal limit.

Then, it is natural to ask how much our results including only thermal quarks can be affected by
thermal hadronic excitations at finite density.
In the crossover regime, which we are most interested in, there can be thermal hadronic excitations.

Computing hadron spectra in the pion condensed phase
requires multi-channel coupling analyses,
as the condensed pions serve isospin and change the parity when other particles propagate.
Three pion modes all mix and even sigma mode couples due to the violation of the parity in medium.
Because of these complications, 
in this paper we do not directly calculate the spectra within our model
but simply quote the results from the EFT of the $I_3$ symmetry breaking
to get rough estimates.

To avoid artificial jumps in thermodynamics from the hadronic phase
to the pion condensed phase, 
we keep track of adiabatic changes from three pions in the hadronic phase
to three modes in the pion condensed phase. 
The other possible hadronic excitations are neglected, consistently in the normal and pion condensed phases.
The spectra are continuous because the phase transition is the second order in mean field theories.
Among the three modes in the pion condensed phase,
one mode is massless while the other two modes are massive.

The spectra in the normal hadronic phase are
\beq
&&E_{\pi_\pm} = \sqrt{ m_\pi^2 + \vp^2 \,} \mp 2\mu_I \,, \notag \\
&&E_{\pi_0} = \sqrt{ m_\pi^2 + \vp^2 \,}  \,.
\eeq
The $\pi_+$ becomes massless at $\mu_I = m_\pi/2$.

In the pion condensed phase, we use the labels $\tilde{\pi}_i$ which are smoothly connected with $\pi_i$.
The spectra are (see Appendix~\ref{sec:pion_spectra})
\beq
&& E_{\tilde{\pi}_+ } =  |\vp| \sqrt{ \frac{\, 4 \mu_I^2 -  m_\pi^2 \,}{\, 12 \mu_I^2 -  m_\pi^2 \,}  \,} 
\equiv v_\pi |\vp| \,, \notag \\
&& E_{\tilde{\pi}_-} = \sqrt{ m_{-}^2 + \bigg( 1 + \frac{ 16\mu_I^2 }{\,  m_{-}^2 \,}\bigg) \vp^2  \,} \,, \notag \\
&& E_{\tilde{\pi}_0} = \sqrt{ 4\mu_I^2 + \vp^2 \,} \,,  
\eeq
where $ m_{-}^2 = 24 \mu_I^2 - 2m_\pi^2$ whose derivation is model dependent.
One can check the spectra at $\mu_I=m_\pi/2$ coincides with the spectra in the normal phase.

We note that at large $\mu_I$, the velocity of the massless mode approaches $\sqrt{1/3}$.
Meanwhile, at the onset of the pion condensation, $\Delta \rightarrow 0$,
the velocity $v_\pi$ of $\tilde{\pi}_+$ vanishes.
As shown in the Appendix~\ref{sec:pion_spectra},
the velocity scales as $v_\pi \sim \Delta$.
This is the critical slowing down.

When we try to include the massless mode in the thermodynamics,
some additional considerations are necessary for the applicability of the EFT.
Formally we find
\beq
P_{\tilde{\pi}_+}
= - T \int_{\vp} \ln \big( 1 - \rme^{- \beta v_\pi |\vp| } \big) \,.
\eeq
If we take this expression at its face value,
we run into troubles near the threshold;
$P_{\tilde{\pi}_+}$ diverges as $ \big( T/v_{\pi} \big)^3 \rightarrow \infty$
for $\mu_I \rightarrow m_\pi/2 +0^+$.
This divergence is fictitious because
the phase space where the EFT is applicable is limited to $\sim \Delta $.
Including such cutoff scale in the momentum integral,
the change of variable as $\sim v_\pi \vp \rightarrow \vp'$
yields the pressure which parametrically behaves as
\beq
P_{\tilde{\pi}_+} \sim T \Delta^3 \,,
\eeq
and is harmless for $\mu_I \rightarrow m_\pi/2 +0^+$ or $\Delta \rightarrow 0$.
In practice, we integrate momenta of $\tilde{\pi}_+$ up to $2\Delta$
beyond which pions may decay to a quark and a quark-hole.

\begin{figure}[tbp]
\vspace{-0.cm}
\centering
\includegraphics[width=1.0\linewidth]{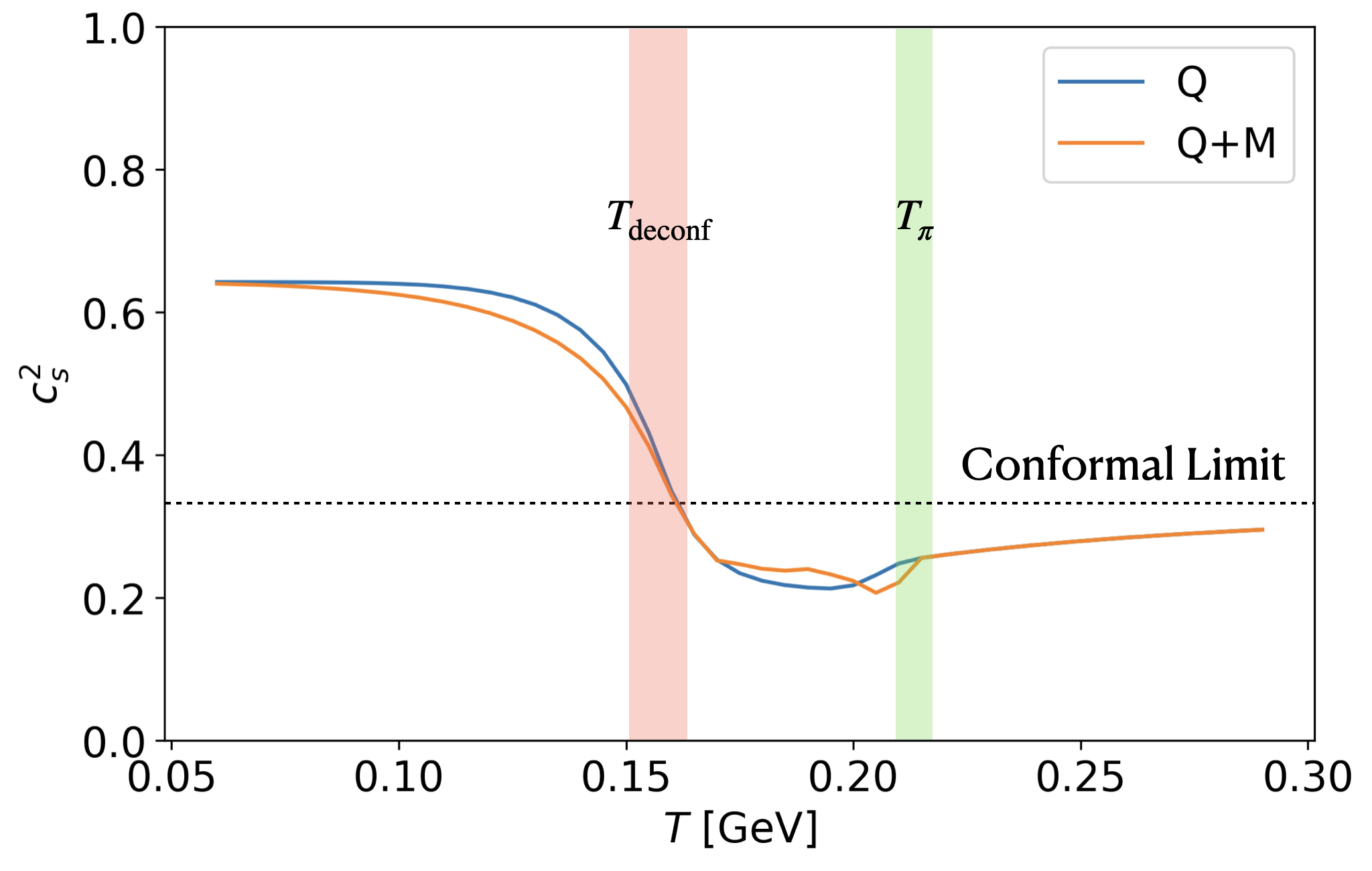}
\caption{ The squared sound velocity $c_s^2$ at $\mu_I=0$ and various temperatures
	at $\mu_I = 240$ MeV.
	Results with and without thermal pions are compared.
	The quick reduction of $c_s^2$ occurs in the interval $T=$ 140-170 MeV
	which coincides with the domain where the Polyakov loop rises quickly.
	}
\label{fig:cs_sn_240.png}
\end{figure}

Shown in Fig.~\ref{fig:cs_sn_240.png} is $c_s^2$ at $\mu_I=240$ MeV at various temperatures.
The isospin density is $\sim 9n_0$ which is smaller than the density where pions are supposed to overlap.
Results with and without thermal pions are compared.
To moderate temperature $\sim 100$ MeV, thermal pions give minor corrections,
reflecting that the cold matter pressure at $T=0$ is already substantial.

%

\section{Summary}
\label{sec:summary}

We study the EOS of the isospin QCD within a quark-meson model with the Polyakov loop at finite $\mu_I$ and $T$,
especially along isentropic trajectories where $s/n_I$ is fixed.
The model describes the BEC-BCS crossover of pion condensates.
At low density the condensed pions dominate the EOS while at high density they play roles of the BCS gaps for quarks.
Based on the previous analyses for the sound velocity peak and trace anomaly at $T=0$ \cite{Chiba:2023ftg},
in this work we examine corrections from thermally excited quarks in the Polyakov loop background.
The thermal corrections smear out the sound velocity peak and
eventually bring the trace anomaly to the positive value.
The sound velocity becomes smaller than the conformal value
when thermal quarks, which appear to be non-relativistic, are liberated.
This changes occur within pion condensed phase whose melting temperature is 
greater than the temperature for quark liberation (with substantial Polyakov loops).


In this work we include only thermal quarks but neglected thermally excited dynamical gluons and hadrons.
A part of gluons are included in the form of Polyakov loop potential,
but in a QGP thermal gluons should be manifestly taken into account.
A more qualitatively nontrivial question is the impact of  hadronic excitations in the pion condensed phases.
In this paper we gave only rough estimates in the region close to the quark matter domain
and concluded that thermal hadronic contributions give minor corrections to the zero temperature EOS.
But more quantitative discussions are necessary in the domain closer to the hadronic domain
where zero temperature EOS can be comparable to thermal contributions.
The description requires the understanding of spectra, momentum dependence, and the dissociation scale.
In the pion condensed phase, all pions and $\sigma$ mesons are mixed so that 
the coupled channel computations based on quarks are necessary.
In addition to the impacts on EOS, 
studying the properties of hadronic excitations in dense matter is of interest 
by its own (see Refs.~\cite{Suenaga:2022uqn,Murakami:2022lmq,Suenaga:2023xwa} for two-color QCD),
as they are relevant for transport phenomena.
In future we plan to perform extensive analyses based on quark degrees of freedom.

Another interesting question is how gluons and the Polyokov loop potential are modified by the medium effects.
In two-color QCD, lattice QCD simulations indicate that
modifications to gluon propagators are modest even at quark chemical potential 
of $\mu_q \sim 1$ GeV \cite{Kojo:2014vja,Suenaga:2019jjv,Kojo:2021knn,Contant:2019lwf,Boz:2018crd,Bornyakov:2020kyz}.
The Polyakov loop potential calculated from such gluon propagators should also 
show modest response to changes in $\mu_I$.
Since the non-perturbative power corrections are very important 
for $c_s^2$ to exceed the conformal value $1/3$,
it is crucial to understand how the non-perturbative effects become irrelevant at high density.
Such analyses will allows us to judge which conceptions in QCD-like theories 
can be transferred to QCD at finite $\mu_B$.

\begin{acknowledgments}
We thank Drs. Brandt and Endrodi for kindly providing us with their lattice data in Ref.~\cite{Brandt:2022hwy},
and Dr. Abbott and his collaborators for their kindness of sending the lattice data in Ref.~\cite{Abbott:2023coj}.
We thank Drs. 
E. Itou, 
Y. Fujimoto 
for discussions.
This work was supported by JSPS KAKENHI Grant No. 23K03377.
TK is also supported by the Graduate Program on Physics for the Universe (GPPU) at Tohoku university;
D.S. by the RIKEN special postdoctoral researcher program and the JSPS KAKENHI Grant No. 23H05439.
\end{acknowledgments}

\appendix

\section{Pion spectra in the pion condensed phase}\label{sec:pion_spectra}

We consider an EFT$_{I_3}$ for charged pions,
\begin{align}
\calL & =  (\partial_\mu + 2 \rmi \mu_I\delta_\mu^0) \pi^+\left(\partial^\mu - 2 \rmi \mu_I\delta^\mu_0\right)\pi^- \notag\\
	 & \hspace{0.5cm} - {m_{\pi}^2\over 2} \big( \pi_1^2 + \pi_2^2 \big)  - {\lambda' \over 24} \big( \pi_1^2 + \pi_2^2 \big)^2 + \cdots  
\end{align}
We assume that quark and $\sigma$ are integrated out
and hence $\lambda'$ may be modified from the parameters in the quark-meson model.
Pions here are just treated as ordinary massive mesons with the mass $m_\pi$.
The following calculations do not refer to the chiral symmetry in the underlying theories.
The condensates are given by
\beq
\bar{\pi}^2 
= \frac{\, 6 \,}{\lambda'} \big( 4\mu_I^2 - m_\pi^2 \big) \,.
\eeq
With this background, the fluctuation of pions are given by 
%
\beq
\calL_{1,2} 
&=&  {1\over2}\left[(\partial_\mu \pi_1)^2+(\partial_\mu \pi_{2} )^2\right]
\notag \\
&&- 2\mu_I \left[(\partial_0 \pi_1) \pi_2 - (\partial_0 \pi_2) \pi_1 \right]
- \frac{\, \lambda' \bar{\pi}^2 \,}{\, 6 \,} \pi_1^2 \,.
\eeq
The spectra for $\tilde{\pi}_\pm$ are obtained by finding $p_0$ such that
\beq
\det
\left[
\begin{matrix}
~p_0^2 - \vp^2 +\frac{\, \lambda' \,}{3} \bar{\pi}^2  ~&~ 4\rmi \mu_I p_0 \\
 -4\rmi \mu_I p_0 ~&~ p_0^2 - \vp^2 ~
\end{matrix}
\right]
=0 \,.
\eeq
For $\vp=0$, it is clear $p_0 =0$ satisfies the equation, i.e., there is massless mode.
Meanwhile, a massive mode has the squared mass
\beq
m_-^2 = 16 \mu_I^2 + \frac{\, \lambda' \,}{3} \bar{\pi}^2
= 24 \mu_I^2 - 2m_\pi^2 \,.
\eeq
At finite $\vp$, the spectra are given by
\beq
E_{\tilde{\pi}_-} 
= \sqrt{ m_{-}^2 + \bigg( 1 + \frac{ 16\mu_I^2 }{\,  m_{-}^2 \,}\bigg) \vp^2 }
\eeq
for a massive mode, and
\beq
E_{\tilde{\pi}_+} = v_\pi |\vp| \,,
\eeq
with
\beq
v_\pi 
\equiv 
\sqrt{ \frac{\, 4 \mu_I^2 -  m_\pi^2 \,}{\, 12 \mu_I^2 -  m_\pi^2 \,} } 
= \bar{\pi} \sqrt{ \frac{\, \lambda'  \,}{\, 3  m_{- }^2 \,} } \,.
\eeq
We note that at large $\mu_I$, the velocity of the massless mode approaches $\sqrt{1/3}$.
Meanwhile, at the onset of the pion condensation, $\bar{\pi} \rightarrow 0$,
the velocity $v_\pi$ of $\tilde{\pi}_+$ vanishes.
This is the critical slowing down.



\bibliography{ref}

\end{document}